\documentclass[%
 reprint,
superscriptaddress,
 amsmath,amssymb,
aps,
pra,
]{revtex4-2}

\usepackage{algorithm}
\usepackage{algorithmic}
\usepackage{graphicx}
\usepackage{bm}
\usepackage{amsthm,amsmath,amssymb}
\usepackage{color}
\usepackage{enumitem}
\usepackage{caption}
\usepackage{subcaption}
\usepackage{mathtools}
\usepackage{dcolumn}
\usepackage{hyperref}

\definecolor{Red}{rgb}{1,0,0}

\begin{document}

\preprint{APS/123-QED}

\title{Quantum classifiers for domain adaptation}

\author{Xi He}
\affiliation{Shaanxi Joint Laboratory of Artificial Intelligence, Shaanxi University of Science and Technology, Xi'an 710021, China}
\affiliation{School of Electronic Information and Artifical Intelligence, Shaanxi University of Science and Technology, Xi'an 710021, China}
\affiliation{Institute of Fundamental and Frontier Sciences, University of Electronic Science and Technology of China, Chengdu, 610054, China}

\author{Feiyu Du}
\affiliation{Shaanxi Joint Laboratory of Artificial Intelligence, Shaanxi University of Science and Technology, Xi'an 710021, China}
\affiliation{School of Electronic Information and Artifical Intelligence, Shaanxi University of Science and Technology, Xi'an 710021, China}

\author{Mingyuan Xue}
\affiliation{Shaanxi Joint Laboratory of Artificial Intelligence, Shaanxi University of Science and Technology, Xi'an 710021, China}
\affiliation{School of Electronic Information and Artifical Intelligence, Shaanxi University of Science and Technology, Xi'an 710021, China}

\author{Xiaogang Du}
\affiliation{Shaanxi Joint Laboratory of Artificial Intelligence, Shaanxi University of Science and Technology, Xi'an 710021, China}
\affiliation{School of Electronic Information and Artifical Intelligence, Shaanxi University of Science and Technology, Xi'an 710021, China}

\author{Tao Lei}
\email{leitao@sust.edu.cn}
\affiliation{Shaanxi Joint Laboratory of Artificial Intelligence, Shaanxi University of Science and Technology, Xi'an 710021, China}
\affiliation{School of Electronic Information and Artifical Intelligence, Shaanxi University of Science and Technology, Xi'an 710021, China}

\author{A. K. Nandi}
\affiliation{Department of Electronic and Electrical Engineering, Brunel University London, Uxbridge, Middlesex, UB8 3PH, U.K.}
\affiliation{School of Mechanical Engineering, Xi'an Jiaotong University, Xi'an 710049, China}

\begin{abstract}
Transfer learning (TL), a crucial subfield of machine learning, aims to accomplish a task in the target domain with the acquired knowledge of the source domain. Specifically, effective domain adaptation (DA) facilitates the delivery of the TL task where all the data samples of the two domains are distributed in the same feature space. In this paper, two quantum implementations of the DA classifier are presented with quantum speedup compared with the classical DA classifier. One implementation, the quantum basic linear algebra subroutines (QBLAS)-based classifier, can predict the labels of the target domain data with logarithmic resources in the number and dimension of the given data. The other implementation efficiently accomplishes the DA task through a variational hybrid quantum-classical procedure. 

\end{abstract}




\maketitle
\section{Introduction}
\label{sec:introduction}
Transfer learning (TL), a significant subfield of machine learning, attempts to accomplish tasks in an unknown domain with the learnt knowledge of a different, but related domain~\cite{1}. As a crucial sub-realm of the TL, domain adaptation (DA) aims to predict the labels of an unlabelled target domain with a given labelled source domain data where all the given data are distributed in the same feature space. DA is significantly applicable in dealing with the unprocessed data and has been widely used in various fields such as computer vision~\cite{2,3,4}, natural language processing~\cite{5}, and reinforcement learning~\cite{6}.

Quantum computing is a type of pattern computing which is typically based on quantum mechanics~\cite{7, 8, 9, 10, 11}. In recent years, quantum computing techniques have been applied to the field of machine learning to accomplish tasks with the promotion of the algorithm performance~\cite{12, 13, 14}. For instance, quantum computation techniques can achieve supervised learning tasks such as classification~\cite{15,16,17}, data fitting~\cite{18,19}, and unsupervised learning such as clustering~\cite{20}, dimensionality reduction~\cite{21,22} with quantum speedup. In the field of deep learning, quantum Boltzmann machine~\cite{23,24}, quantum generative adversarial learning~\cite{25,26,27,28,29,30}, quantum auto-encoder~\cite{31,32,33}, and quantum neural networks~\cite{34,35} have been proposed to deal efficiently with deep learning tasks on quantum devices. For the TL, Ref.~\cite{36} systematically analyzes the framework of the quantum transfer learning in different scenarios. Ref.~\cite{37,38} utilize linear transformation to align the source domain to the target domain to accomplish the procedure of DA. However, the procedure of the DA and the labels prediction are specifically separated in the existing quantum DA algorithms resulting in an increase of the computational complexity.  

In this paper, two quantum implementations of the DA classifier are presented. One implementation utilizes the quantum basic linear algebra subroutines to achieve exponential speedup on the universal quantum computer compared to the classical DA classification algorithm. The other implementation, the variational quantum DA classifier, accomplishes the procedure of DA on the near-term quantum devices through a variational hybrid quantum-classical procedure. 

The remainder of this paper is arranged as follows. Firstly, the classical DA classifier is briefly overviewed in section~\ref{sec:classical DAC}. Subsequently, the QBLAS-based and the variational quantum DA classifiers are presented respectively. Finally, some open problems and future work are discussed. 

\section{Classical domain adaptation classifier}
\label{sec:classical DAC}
Assume that we are given a source domain dataset $\mathcal{D}_{s} = \{ x_{i}^{(s)} \}_{i=1}^{n_{s}} \in \mathbb{R}^{D}$ with labels $\{ y_{i}^{(s)} \}_{i=1}^{n_{s}} \in \{ 0, 1 \}$ and an unlabelled target domain dataset $\mathcal{D}_{t} = \{ x_{j}^{(t)} \}_{j=1}^{n_{t}} \in \mathbb{R}^{D}$. The source domain data matrix $X_{s} = (x_{1}^{(s)}, \cdots, x_{n_{s}}^{(s)}) \in \mathbb{R}^{D \times n_{s}}$ and the target domain data matrix $X_{t} = (x_{1}^{(t)}, \cdots, x_{n_{t}}^{(t)}) \in \mathbb{R}^{D \times n_{t}}$. The feature and the label space of $\mathcal{D}_{t}$ are exactly the same as $\mathcal{D}_{s}$. However, the data of the source and target domain specifically obey different data distributions. The goal of the classifier for domain adaptation is to predict the labels of an unknown target domain with the help of the labelled source domain data~\cite{39}. Let $\mu_{c}^{(s)}$ $(\mu_{c}^{(t)})$, $\Sigma_{s}$ $(\Sigma_{t})$ be the $c$th class mean and the covariance of the source (target) domain for $c \geq 2$. In this paper, the discussion is specifically restricted to the task of binary classification, namely $c=2$. The binary domain adaptation classifier can be easily extended to the circumstance of multi-class.   

\begin{figure}[h]
\centering
\includegraphics[width=0.45\textwidth]{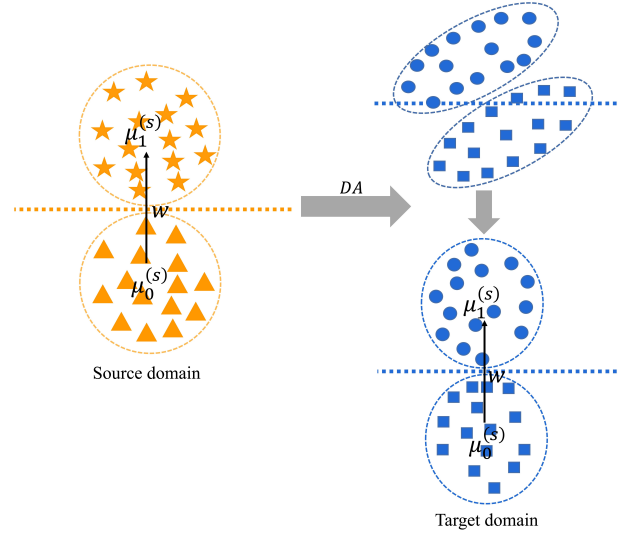}
\caption{The schematic diagram of the DA classifier. The covariance matrix $\Sigma_{s}$ is applied to the mean of the data feature without background class resulting in the weight of the DA classifier. Combined with the decorrelated target domain data, the DA classifier can be constructed.}
\label{fig:DA classifier}
\end{figure}

The DA classifier achieves the procedure of the transfer learning with a modified classifier inspired from the linear discriminant analysis (LDA)~\cite{40}. The scoring function of the classifier is defined as
\begin{equation}
	y(x) = w^{T} x
	\label{eq:scoring function}
\end{equation}
to determine the label of the specified data point. In the spirit of the LDA, the source domain data $\mathcal{D}_{s}$ are generated from the distribution $p(x^{(s)}, y^{(s)}) = p(x^{(s)} | y^{(s)})p(y^{(s)})$ where $p(y^{(s)})$ is the prior of the labels; $p(x^{(s)} | y^{(s)}) = \mathcal{N}(x^{(s)}; \mu_{c}^{(s)}, \Sigma_{s})$ represents the class-conditional distributions. The weight vector of the classifier is $w^{(s)} = \Sigma_{s}^{-1/2} (\mu_{1}^{(s)} - \mu_{0}^{(s)})$. Equivalently, the classifier can be obtained by projecting the decorrelated source domain data $\hat{x} = \Sigma_{s}^{-1/2} x^{(s)}$ to the difference between the decorrelated means $\hat{w} = (\hat{\mu_{1}}^{(s)} - \hat{\mu_{0}}^{(s)}) = \Sigma_{s}^{-1/2}(\mu_{1}^{(s)} - \mu_{0}^{(s)})$ where $\hat{u}_{c}^{(s)} = \Sigma_{s}^{-1/2} \mu_{c}$ for $c = 0, 1$. However, this classifier cannot be directly applied to the target domain directly due to the domain shift between the source and target domain. The DA classifier adaptively apply the decorrelated target domain data $\hat{x}^{(t)} = \Sigma_{t}^{-1/2} x^{(t)}$ to the decorrelated mean difference $\hat{w}$ resulting in the scoring function
\begin{equation}
	\begin{split}
		\hat{y}(x^{(t)}) &= \hat{w}^{T} \hat{x}_{t} \\
		&= (\Sigma_{s}^{-1/2}(\mu_{1}^{(s)} - \mu_{0}^{(s)}))^{T} (\Sigma_{t}^{-1/2} x^{(t)}).
	\end{split}
	\label{eq:DAC scoring function}
\end{equation}
The DA classifier utilizes the DA techniques to modify the traditional model for classification to effectively accomplish the machine learning tasks in different domains. Compared with other DA models, the DA classifier effectively combines the procedure of TL with the label prediction resulting in a concise DA model. The schematic diagram of the DA classifier is presented as~\ref{fig:DA classifier}.  

\section{State preparation}
\label{sec:state preparation}
Given the source domain data $X_{s}$ and the target domain data $X_{t}$, the quantum states corresponding to the $X_{s}$ and $X_{t}$ are 
\begin{equation}
	| \psi_{X_{s}} \rangle = \sum_{i=1}^{n_{s}}\sum_{m=1}^{D} x_{mi}^{(s)} | i \rangle | m \rangle = \sum_{i=1}^{n_{s}} | i \rangle | x_{i}^{(s)} \rangle, 
	\label{eq:psi_X_s}
\end{equation}
\begin{equation}
	| \psi_{X_{t}} \rangle = \sum_{j=1}^{n_{t}}\sum_{m=1}^{D} x_{mi}^{(t)} | j \rangle | m \rangle = \sum_{j=1}^{n_{t}} | j \rangle | x_{j}^{(t)} \rangle, 
	\label{eq:psi_X_t}
\end{equation}
respectively in amplitude encoding where $\sum_{m, i} | x_{mi}^{(s)}|^{2} = \sum_{mj} | x_{mj}^{(t)} |^{2} = 1$~\cite{41}. Hence, the states which represent the covariance matrices of the source and target domain data are 
\begin{equation}
		\rho_{s} = \mathrm{tr}_{i} \{ | \psi_{X_{s}} \rangle \langle \psi_{X_{s}} | \} = \sum_{m, m^{'}=1}^{D} \sum_{i=1}^{n_{s}} x_{mi}^{(s)} x_{m^{'}i}^{(s)\ast} | m \rangle \langle m^{'} |,  
	\label{eq:source domain covariance state}
\end{equation} 
\begin{equation}
	\rho_{t} = \mathrm{tr}_{j} \{ | \psi_{X_{t}} \rangle \langle \psi_{X_{t}} | \} = \sum_{m, m^{'}=1}^{D} \sum_{j=1}^{n_{t}} x_{mj}^{(t)} x_{m^{'}j}^{(t)\ast} | m \rangle \langle m^{'} |,  
\label{eq:target domain covariance state}
\end{equation} 
respectively where $\mathrm{tr}_{i}$ is the trace over the $i$ register. The quantum states $| \mu_{c}^{(s)} \rangle$ $(c = 0, 1)$ representing the source domain mean value for the two classes can be obtained by the quantum adder proposed in Ref.~\cite{42,43,44}. The quantum states $| \mu_{1}^{(s)} - \mu_{0}^{(s)} \rangle$ can be computed by the quantum subtractor presented in Ref.~\cite{22}. In addition, the data matrices $X_{s}$, $X_{t}$ can be extended to $\tilde{X}_{s} = | 0 \rangle \langle 1 | \otimes X_{s} + | 1 \rangle \langle 0 | \otimes X_{s}^{\dagger}$ and $\tilde{X}_{t} = | 0 \rangle \langle 1 | \otimes X_{t} + | 1 \rangle \langle 0 | \otimes X_{t}^{\dagger}$ respectively.

\section{QBLAS-based DA classifier}
\label{sec:QBLAS-based DAC}
\begin{figure}[h]
\centering
\includegraphics[width=0.45\textwidth]{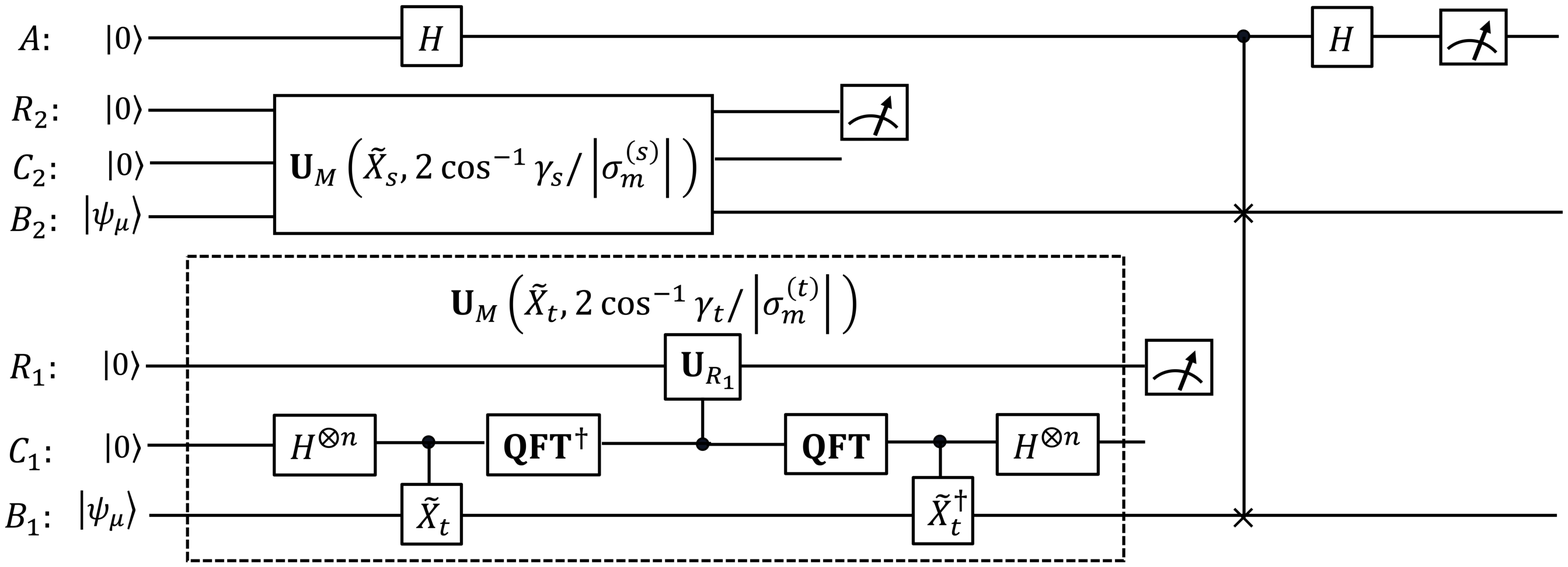}
\caption{The quantum circuits of the QBLAS-based DA classifier.}
\label{fig:QBLAS-based DA classifier}
\end{figure}

For the QBLAS-based DA classifier, we assume that the elements of $X_{s}$ and $X_{t}$ are accessible by the quantum random access memory (qRAM)~\cite{45} in time $O(\mathrm{poly}(\mathrm{log}(D \tilde{n})))$ with $O(\mathrm{poly} (D \tilde{n}))$ resources where $\tilde{n} = \max(n_{s}, n_{t})$. The corresponding quantum circuit of the QBLAS-based DA classifier is depicted as~\ref{fig:QBLAS-based DA classifier}.

The whole procedure of the QBLAS-based DA classifier is presented as follows.

(1) Apply the quantum phase estimation algorithm (QPE)~\cite{46,47}
\begin{equation}
	\textbf{U}_{PE}(\tilde{X}_{t}) = (\textbf{QFT}^{\dagger} \otimes \textbf{I}) \left ( \sum_{\tau=0}^{T-1} | \tau \rangle \langle \tau | \otimes e^{i \tilde{X}_{t} \tau t / T} \right ) (\textbf{H}^{\otimes n} \otimes \textbf{I})
	\label{eq:PE}
\end{equation}
on the input state $| 0\rangle | \psi_{X_{t}} \rangle | 0 \rangle^{\otimes \mathrm{log}(D + n_{t})}$ prepared by the specified registers resulting in the quantum state
\begin{align}
	| \psi_{1} \rangle &= \sum_{j=1}^{n_{t}} | j \rangle \sum_{m=1}^{D} \beta_{mj}^{(t)} | \sigma_{m}^{(t)} \rangle \frac{1}{\sqrt{2}} (| w_{m}^{(t)+} \rangle - | w_{m}^{(t)-} \rangle) \notag \\
	&= | 1 \rangle \sum_{j=1}^{n_{t}} | j \rangle \sum_{m=1}^{D} \beta_{mj}^{(t)} | \sigma_{m}^{(t)} \rangle | v_{m}^{(t)} \rangle, 
	\label{eq:psi_1}
\end{align}
where $\textbf{QFT}^{\dagger}$ represents the inverse quantum Fourier transform~\cite{48}, $\beta_{mi}^{(t)} = \langle u_{m}^{(t)} | x_{j}^{(t)} \rangle$, $| w_{m}^{(t)\pm} \rangle = \frac{1}{\sqrt{2}} (| 0 \rangle | u_{m}^{(t)} \rangle \pm | 1 \rangle | v_{m}^{(t)} \rangle)$ are the eigenvectors of $\tilde{X}_{t}$ corresponding to the singular values $\sigma_{m}^{(t)}$.

(2) Perform the controlled $R_{y}(2 \arccos(\gamma_{t} / \vert \sigma_{m}^{(t)} \vert))$ operation on the first register to obtain the state
\begin{equation}
	| \psi_{2} \rangle = | \psi_{a}^{(t)} \rangle \sum_{j=1}^{n_{t}} | j \rangle \sum_{m=1}^{D} \beta_{mj}^{(t)} | \sigma_{m}^{(t)} \rangle | v_{m}^{(t)} \rangle, 
	\label{eq:psi_2}
\end{equation}
where 
\begin{equation}
	| \psi_{a}^{(t)} \rangle = \sqrt{1 - \frac{\gamma_{t}^{2}}{\vert \sigma_{m}^{(t)} \vert^{2}}} | 0 \rangle + \frac{\gamma_{t}}{\sigma_{m}^{(t)}} | 1 \rangle,
	\label{eq:ancilla}
\end{equation}
$\gamma_{s}$ is a constant.

(3) Uncompute the $| \sigma_{m}^{(t)} \rangle$ register, remove the ancilla register, and measure the $| \psi_{a}^{(t)} \rangle$ to be $| 1 \rangle$. The state
\begin{align}
	| \psi_{\tilde{X}_{t}} \rangle &= \sum_{j=1}^{n_{t}} | j \rangle \sqrt{\frac{1}{\sum_{m=1}^{D} | \gamma_{t} \beta_{mj}^{t} |^{2} / | \sigma_{m}^{(t)} |^{2}}} \sum_{m=1}^{D} \frac{\beta_{mj}^{(s)} \gamma_{t}}{| \sigma_{m}^{(t)} |} | v_{m}^{(t)} \rangle \notag \\
	&= \sum_{j=1}^{n_{t}} | j \rangle \frac{\Sigma_{t}^{-1} | x_{j}^{(t)} \rangle}{\sqrt{\langle x_{j}^{(t)} | \Sigma_{(t)}^{-1 \dagger} \Sigma_{(t)}^{-1} | x_{j}^{(t)} \rangle}} \notag \\
	&= \sum_{j=1}^{n_{t}} | j \rangle | \hat{x}_{j}^{(t)} \rangle
	\label{eq:psi_tilde_X_t}
\end{align}
can be obtained. Thus, the quantum state $| \hat{x}^{(t)}_{j} \rangle$ proportional to $\Sigma_{t}^{- \frac{1}{2}} x^{(t)}_{j}$ for $j = 1, 2, \cdots, n_{t}$ can be computed in time $O(\Vert X_{t} \Vert^{2}_{\max} \log^{2}(D + n_{s}) / \epsilon^{3})$ where $\Vert X_{t} \Vert_{\max}$ is the largest absolute element of $X_{t}$ and $\epsilon$ is the error parameter. 

The whole procedure above can be represented as the following unitary operation
\begin{equation}
	\textbf{U}_{M}(X, \theta) = (\textbf{I} \otimes \textbf{U}_{PE}^{\dagger}(X))(\textbf{U}_{R_{1}}(\theta) \otimes \textbf{I})(\textbf{I} \otimes \textbf{U}_{PE}(X))
	\label{eq:U_M}
\end{equation}
where $| \psi_{\tilde{X}_{t}} \rangle$ can be achieved by the operation $\textbf{U}_{M}(\tilde{X}_{t}, 2\arcsin(\gamma_{t} / \vert \sigma_{m}^{(t)} \vert))$. Similarly, the quantum state $| \psi_{\mu} \rangle$ proportional to the vector $\Sigma_{s}^{-1/2}(\mu_{1}^{(s)} - \mu_{0}^{(s)})$ by applying $\textbf{U}_{M}(\tilde{X}_{s}, 2\arcsin(\gamma_{s} / \vert \sigma_{m}^{(s)} \vert))$ on the input quantum state $| 0 \rangle^{R} | 0\rangle^{C} | \psi_{\mu} \rangle^{B} (| 0 \rangle^{\otimes \mathrm{log}(D + n_{t})})^{S}$ in time $O(\Vert X_{s} \Vert^{2}_{\max} \log^{2}(D + n_{s}) / \epsilon^{3})$~\cite{49} where $\Vert X_{s} \Vert_{\max}$ is the largest absolute element of $X_{s}$. 

(4) The scoring function of the QBLAS-based DA classifier can be ultimately obtained as 
\begin{equation}
	\hat{y}_{q}(x^{(t)}) = \langle \hat{w}_{q} | \hat{x}^{(t)} \rangle 
	\label{eq:QFDA_scoring_function}
\end{equation}
by performing the swap test~\cite{50} on $| \hat{w} \rangle$, $| \hat{x}_{t} \rangle$.

The pseudo-code of the QBLAS-based DA classifier is presented in~\ref{alg:QBLAS-based DA classifier}.

\begin{algorithm}[H]
	\caption{Quantum fast domain adaptation}
	\begin{algorithmic}[]
		\STATE
		\textbf{Input:} Source domain data $X_{s}$ with labels $Y_{s}$, target domain data $X_{t}$. \\
		\textbf{Output:} Target domain labels $Y_{t}$. \\
		\textbf{Step 1:} Perform $\textbf{U}_{PE}(\tilde{X}_{t})$ on $| 0 \rangle | 0\rangle | \psi_{X_{t}} \rangle | 0 \rangle^{\otimes \mathrm{log}(D + n_{t})}$ to obtain $| \psi_{1} \rangle$. \\
		\textbf{Step 2:} Perform the controlled $R_{y}(2 \arcsin(\gamma_{t} / \vert \sigma_{m}^{(t)} \vert))$ rotation operation on $| \psi_{1} \rangle$ to compute $| \psi_{2} \rangle$. \\ 
		\textbf{Step 3:} Uncompute the $| \sigma_{m}^{(t)} \rangle$ register, remove the ancilla register, and measure the $| \psi_{a}^{(t)} \rangle$ to be $| 1 \rangle$ to achieve $| \psi_{\tilde{X}_{t}} \rangle$. \\
        \textbf{Step 4:} Perform the swap test on the weight $| \hat{w}_{q} \rangle$ and decorrelated target domain data $| \hat{x}^{(t)} \rangle$ to predict the target domain labels $\hat{y}_{q}$.
	\end{algorithmic}
	\label{alg:QBLAS-based DA classifier}
\end{algorithm}

\section{Variational quantum domain adaptation classifier}
\label{sec:VQDAC}
In addition to the design based on quantum basic linear algebra subroutines, the DA classifier can be alternatively implemented on noisy intermediate-scale quantum devices (NISQ) through a variational hybrid quantum-classical procedure. The variational quantum domain adaptation classifier (VQDAC) can be performed on the near-term quantum devices without high-depth quantum circuits and fully coherent evolution required by the QBLAS-based DA classifier. The pseudo-code of the VQDAC is presented in~\ref{alg:VQDAC}.

\begin{algorithm}[H]
	\caption{VQDAC}
	\begin{algorithmic}
	\STATE 
	\textbf{Input:} Source domain data $X_{s}$ with labels $Y_{s}$, target domain data $X_{t}$. \\
	\textbf{Output:} Target domain labels $Y_{t}$. \\
	\textbf{Step 1:} Prepare the quantum states $\rho_{s}$, $\rho_{t}$ by the low-depth quantum circuits. \\
	\textbf{Step 2:} Diagonalize $\rho_{s}$, $\rho_{t}$ to construct $\Sigma_{s}^{\frac{1}{2}}$, $\Sigma_{t}^{\frac{1}{2}}$ respectively. \\
	\textbf{Step 3:} Invoke the variational quantum linear solver to compute the decorrelated target domain data $| \hat{x}^{(t)} \rangle$ and the weight coefficient $| w \rangle$. \\
	\textbf{Step 4:} Perform swap test on $| w \rangle$ and $| \hat{x}^{(t)} \rangle$ to predict the target labels $y^{(t)}$.
	\end{algorithmic}
	\label{alg:VQDAC}
\end{algorithm}

For classical data points, the quantum states required can be generated by a quantum circuit of $O(n^{2} + \frac{2^{n}}{k+n})$ depth with $k$ ancilla qubits~\cite{51}. Based on the time-space tradeoff, the quantum states corresponding to the given data can be obtained by low-depth quantum circuits with sufficient quantum qubits. If we are given quantum data initially, the VQDAC can be invoked directly as follows.   

\begin{figure}[h]
\centering
\includegraphics[width=0.45\textwidth]{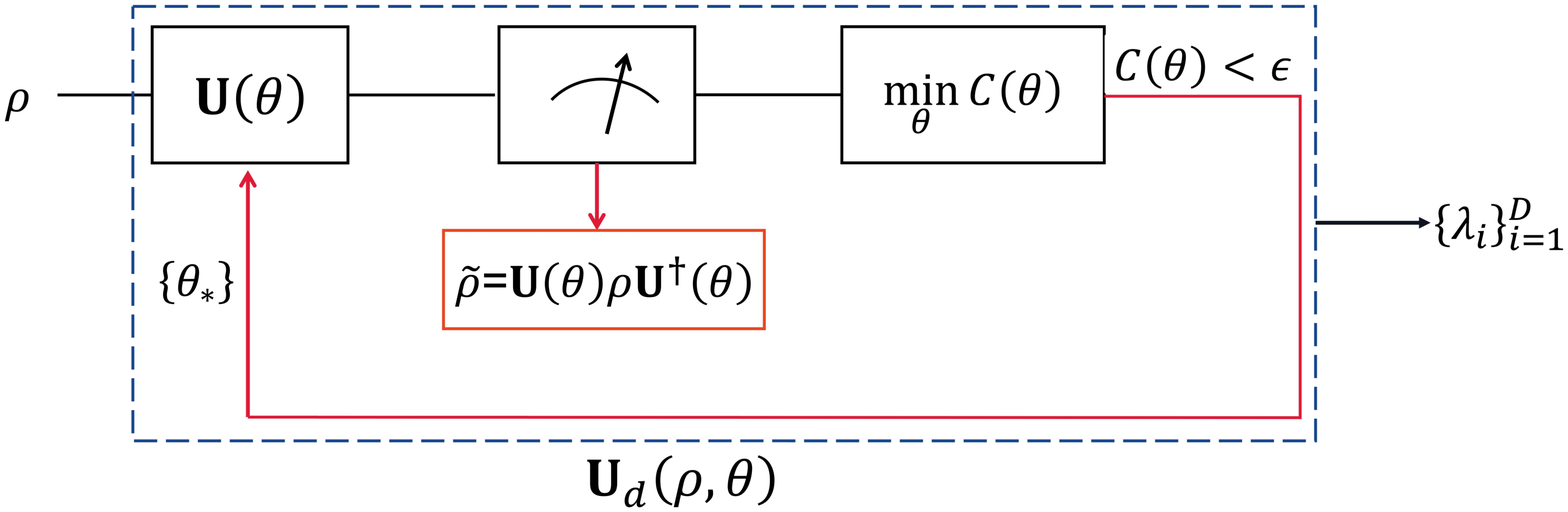}
\caption{The circuits of the diagonalization of a given state $\rho$ through a variational hybrid quantum-classical procedure.}
\label{fig:VQSD}
\end{figure}

The VQDAC firstly diagonalizes the quantum states $\rho_{s}$ and $\rho_{t}$ to obtain the states $| \psi_{s} \rangle$ and $| \psi_{t} \rangle$ to represent the matrix $\Sigma_{s}$ and $\Sigma_{t}$ respectively. In the spirit of Ref.~\cite{52}, design $\tilde{\rho}_{s} = \textbf{U}(\theta_{s}) \rho_{s} \textbf{U}^{\dagger}(\theta_{s})$ with the unitary operation $\textbf{U}(\theta_{s})$ constructed by a parameterized quantum circuit where $\{ \theta_{s} \}$ is a set of parameters. The cost function is defined as 
\begin{equation}
	C = \mathrm{Tr}(\tilde{\rho_{s}} H_{s}),
	\label{eq:cost_function}
\end{equation} 
where $H_{s}$ is a specified $D$-qubit Hamiltonian with $D$ non-negative and non-degenerate eigenvalues. By minimizing the cost function $C$ with a classical optimization algorithm, the optimal parameters $\{ \theta_{s}^{\ast} \}$ can be obtained. $\rho_{s}$'s eigenvalues $\{ \lambda_{i}^{(s)} \}_{i=1}^{D}$ can be estimated by measuring $\tilde{\rho}_{s}$. Thus, the source domain covariance matrix $\Sigma_{s}^{\frac{1}{2}} = \sum_{i=1}^{D} \lambda_{i}^{(s) \frac{1}{2}} | i \rangle \langle i |$ can be finally computed, along with the target domain covariance matrix $\Sigma_{t}^{\frac{1}{2}} = \sum_{j=1}^{D} \lambda_{j}^{(t) \frac{1}{2}} | j \rangle \langle j |$. The quantum circuit of the state diagonalization of the source and target data is presented in~\ref{fig:VQSD}.

Subsequently, the quantum states $| w \rangle$ and $| \hat{x}^{(t)} \rangle$ are computed to represent the weight vector $\hat{w}$ and the target domain whiten data $\hat{X}^{(t)} = \Sigma_{t}^{-\frac{1}{2}} X_{t}$ respectively. Inspired from Ref.~\cite{53}, design the quantum ansatz $| \hat{x}^{(t)} (\theta^{(t)}) \rangle$ with a set of parameters $\{ \theta^{(t)} \}$. The cost function
\begin{equation}
	\mathcal{L} = 1 - \frac{1}{n_{t}} \sum_{j=1}^{n_{t}} \left \vert \frac{\langle x_{j}^{(t)} | \Sigma_{t}^{\frac{1}{2}} | \hat{x}^{(t)} \rangle}{\sqrt{\langle \hat{x}^{(t)}(\theta^{(t)}) | \Sigma_{t}^{\frac{1}{2}\dagger} \Sigma_{t}^{\frac{1}{2}} | \hat{x}^{(t)} (\theta^{(t)}) \rangle}} \right \vert 
	\label{eq:L_1}
\end{equation} 
is defined to be minimized by the classical optimization algorithm such as stochastic gradient descent to obtain the optimal coefficients $\{ \theta^{(t)}_{\ast} \}$ and the decorrelated target domain data $| \hat{x}_{j}^{(t)} (\theta_{\ast}^{(t)}) \rangle$ for $j = 1, \cdots, n_{t}$ in time $O(\kappa_{t} / \epsilon)$, where $\kappa_{t}$ is the condition number of $\Sigma_{t}$ and $\epsilon$ is the error coefficient. Similarly, the quantum state $| w \rangle$ which represents the weight of the DAC $w = \Sigma_{s}^{-1/2}(\mu_{1}^{(s)} - \mu_{0}^{(s)})$ can be computed in the runtime $O(\kappa_{s} / \epsilon)$ where $\kappa_{s}$ is the condition number of the source domain covariance matrix $\Sigma_{s}$.  

\begin{figure}[h]
\centering
\includegraphics[width=0.45\textwidth]{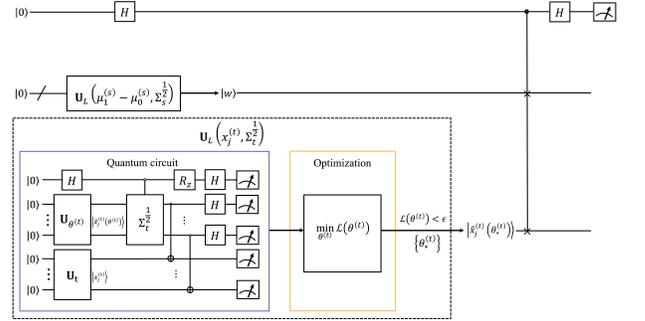}
\caption{The quantum circuits of the VQDAC.}
\label{fig:VQDA}
\end{figure}

Ultimately, the label of the target domain data point $x^{(t)}$ can be obtained according to the success probability of performing the swap test on the two states $| w \rangle$ and $| \hat{x}^{(t)} \rangle$. The whole procedure of the VQDAC is depicted in~\ref{fig:VQDA}.

\section{Conclusions and discussions}
\label{sec:conclusions & discussions}
In this paper, two quantum implementations of the DAC are presented. The QBLAS-based DAC can be implemented on a universal quantum computer with logarithmic resources in the dimension and number of given data. The VQDAC can be performed on the near-term quantum devices through a variational hybrid quantum-classical procedure. 

However, some open questions of the two quantum algorithms need further study. At first, the QBLAS-based DAC requires high-depth quantum circuits and fully coherent evolution in practice. Although it can be proved that the QBLAS-based DAC can achieve quantum speedup, the implementation requirement in practice is relatively hard at present. In addition, the optimal performance of the VQDAC still needs exploration. The specific design of the parameterized quantum circuits is vital to the accuracy of the variational algorithm. How to find the optimal circuit structure is another crucial open question. In spite of the open questions above, it is demonstrated that quantum techniques can be applied to the field of domain adaptation resulting in performance promotion.  

\begin{acknowledgements}
	The author would like to thank Xiaoting Wang for constructive discussions. The author also would like to thank the referees for helpful comments on this paper. This work is supported by National Key Research and Development Program of China Grant No. 2018YFA0306703, in part by the National Natural Science Foundation of China under Grant 61871259, Grant 61861024, in part by Natural Science Basic Research Program of Shaanxi (No. 2021JC-47), and in part by Key Research and Development Program of Shaanxi (NO. 2021ZDLGY08-07)
\end{acknowledgements}

\nocite{*}


\end{document}